\documentclass[preprint,showpacs,amsmath,amssymb,aps,prc,superscriptaddress]{revtex4}
\usepackage{epsfig}
\begin{document}

\preprint{APS/123-QED}

\title{Coupled-channels effects in elastic scattering and near-barrier fusion  
induced by weakly bound nuclei and exotic halo nuclei}

\author{C. Beck}
\thanks{Corresponding author: christian.beck@ires.in2p3.fr}
\affiliation{Institut Pluridisciplinaire Hubert Curien, UMR7178,
IN2P3-CNRS et Universit\'e Louis Pasteur (Strasbourg I), 23 rue du Loess -
BP28, F-67037 Strasbourg Cedex 2, France}
\author{N. Keeley}
\altaffiliation[Permanent address: ]{Department of Nuclear Reactions, The 
Andrzej So\l tan Institute for Nuclear Studies, Ho\.za 69, 
PL-00681, Warsaw, Poland}
\affiliation{DSM/DAPNIA/SPhN CEA Saclay, Orme des
Merisiers, F-91191 Gif-sur-Yvette Cedex, France}
\author{A. Diaz-Torres}
\affiliation{Department of Nuclear Physics, Research School of Physical 
Sciences and Engineering, The Australian National University, Canberra
ACT 0200, Australia}

\date{\today} 

\begin{abstract} 

{The influence on fusion of coupling to the breakup process is investigated for 
reactions where at least one of the colliding nuclei has a sufficiently low 
binding energy for breakup to become an important process. Elastic scattering, 
excitation functions for sub-and near-barrier fusion cross sections, and 
breakup yields are analyzed for $^{6,7}$Li+$^{59}$Co. Continuum-Discretized 
Coupled-Channels (CDCC) calculations describe well the data at and above the 
barrier. Elastic scattering with $^{6}$Li (as compared to $^{7}$Li) indicates 
the significant role of breakup for weakly bound projectiles. A study of 
$^{4,6}$He induced fusion reactions with a three-body CDCC method for the 
$^6$He halo nucleus is presented. The relative importance of breakup and 
bound-state structure effects on total fusion is discussed.}

\end{abstract} 


\maketitle

\section{Introduction} 

In reactions induced by light weakly bound nuclei, the influence on the fusion
process of couplings both to collective degrees of freedom and to breakup (BU) 
or transfer (TR) channels is a key point for a deeper understanding of 
few-body systems in quantum dynamics~\cite{Balantekin98,Dasgupta98}. Due to 
the very weak binding energies of halo nuclei, such as $^{6}$He or $^{11}$Be
\cite{Hagino00,Raabe04,Liang06,Canto06,Keeley07}, a diffuse cloud of neutrons should 
lead to enhanced tunneling probabilities below the Coulomb barrier, where the
neutron tail which extends well beyond the compact nuclear core provides a
conduit by which the matter distributions of the target and projectile may
overlap at longer range than for the core. In the vicinity of the Coulomb 
barrier and below, enhanced fusion yields with $^{11}$Be were predicted
\cite{Hagino00} but not confirmed experimentally for
$^{11,10}$Be+$^{209}$Bi reactions \cite{Signorini04}. For $^6$He, there is
some evidence for enhanced fusion probability compared to the $^4$He core
at deep sub-barrier energies in the $^6$He+$^{206}$Pb system \cite{Penion06}
(a same observation has been recently shown for the $^{197}$Au target 
\cite{Penion07}).
A model of ''sequential fusion'' \cite{Zagrebaev03} where the fusion enhancement 
effect was assumed to be due to the gain in energy from a rearrangement of the 
$^6$He valence neutrons (due to the positive Q-values for one- and two-neutron TR) 
was able to predict successfully the data before the experiment. 
However, most other recent experimental 
studies involving $^{6}$He radioactive ion beams (RIB)
\cite{Raabe04,Kolata98,Aguilera00,Bychowski04,Dipietro04,Navin04,DeYoung05,Kolata05}
indicate that the halo of the $^{6}$He nucleus does not enhance the fusion
probability, but illustrate the preponderant role of one- and two-neutron
TR in $^{6}$He induced reactions~\cite{Bychowski04,DeYoung05}. Hence, 
the question of a real new effect with RIBs and with weakly bound stable beams
such as $^{6}$Li, $^{7}$Li and $^{9}$Be remains open \cite{Liang06,Canto06,Keeley07}: 
namely the occurrence of non-conventional transfer/stripping processes with 
large cross sections most likely originating from the small binding energy of 
the projectile as well as the specific role of the BU process have still to 
be clearly determined. More exclusive measurements will be necessary to 
disentangle the different components.

Since coupling between channels is known to enhance the fusion cross
section at sub-barrier energies \cite{Hagino00,Dasso85}, coupled-channels (CC)
effects have often been taken into account in the theoretical description of
the quantum tunneling in fusing systems
\cite{Hagino00,Balantekin98,Dasgupta98,Canto06}. A large number of experimental 
results have been interpreted adequately well within the framework of CC 
calculations \cite{Balantekin98,Dasgupta98,Canto06,Keeley07}. However, in the case 
of loosely bound (and/or halo) systems the situation is more complicated since the 
BU and TR channels may induce strong couplings to an infinite number of unbound 
states in the continuum of the projectile. A possible treatment of the problem 
is to reduce it to a finite number of states. This is often achieved by 
discretizing in energy the continuum of the weakly bound nucleus such that the 
resulting set of coupled equations may be solved in the conventional manner. 
This is the so-called method of Continuum-Discretized Coupled-Channels (CDCC)
\cite{Rawitscher74,Kamimura86,Yahiro86,Sakuragi86,Austern87,Tostevin01,Diaz02,Keeley02,
Rusek03,Keeley03,Diaz03,Rusek04,Mackintosh04,Keeley05,Rusek05}. 
With the recent advent of new RIB facilities~\cite{Liang06,Canto06,Keeley07}, it is 
now necessary to extend the CDCC formalism to allow for four-body BU and/or
excitation of the ``core'' nucleus (the question of the treatment of TR channels
is also still open). Studies have been initiated in this direction by several groups 
\cite{Matsumoto04a,Matsumoto04b,Rodriguez05,Matsumoto06,Summers06a,Summers06b} 
to investigate reactions induced by an exotic ``Borromean" ($^{6}$He) nucleus, 
which is known to have a strong dipole excitation mode \cite{Aumann99}, and the 
single neutron halo nucleus $^{11}$Be, where collective excitation of the 
$^{10}$Be core is expected to be important. 

In this work we present CDCC calculations describing simultaneously the elastic 
scattering and limited available BU data for the weakly bound stable nuclei 
$^{6}$Li and $^{7}$Li interacting with the medium-mass $^{59}$Co target and 
separate calculations for the total fusion (TF) of these nuclei with $^{59}$Co 
and for $^{6}$He with $^{59}$Co and $^{63,65}$Cu. Preliminary reports of this 
work have been presented elsewhere in conference proceedings 
\cite{Beck06a,Beck06b}. A description of the CDCC calculations is given in 
Sec. II. The CDCC results and corresponding comparisons with available 
experimental elastic scattering, BU, and TF cross sections are discussed 
in Sec. III. Section IV provides a brief summary as well as suggesting future 
directions for experimental and theoretical investigations. 

\section{Continuum-discretized Coupled-Channels calculations} 

The fully quantum-mechanical CDCC method, first proposed in the early seventies 
by Rawitscher~\cite{Rawitscher74} to study the effect of deuteron breakup on
elastic scattering, has been widely applied by the Kyushu group
\cite{Kamimura86,Yahiro86,Sakuragi86,Austern87} to study heavy-ion collisions 
induced by light weakly bound nuclei. CDCC calculations have been successful 
in the past in describing the scattering of deuterons~\cite{Yahiro86,Austern87} 
and $^{6,7}$Li~\cite{Sakuragi86} on different targets. The standard three-body 
CDCC method has also been applied to reactions involving halo nuclei, e.g.\ $^8$B 
\cite{Tostevin01} and $^6$He \cite{Rusek03}. Diaz-Torres and Thompson 
\cite{Diaz02} have used a novel method based on the CDCC formalism to perform a 
fully quantum-mechanical calculation of TF of the halo nucleus $^{11}$Be with a 
$^{208}$Pb target, later refined and applied to the TF of $^{6,7}$Li 
\cite{Diaz03}. A recent study of the $^{6}$He+$^{209}$Bi reaction by means of a 
three-body CDCC model~\cite{Keeley03} found much larger absorption cross sections 
than those extracted from optical model (OM) fits to the elastic scattering data 
\cite{Aguilera00}, a problem that is confirmed by more realistic four-body CDCC 
calculations \cite{Matsumoto06} that describe the data well. However, the problem 
with the simple three-body CDCC model for $^6$He breakup has been traced to the 
E1 coupling strength; when these couplings are reduced by 50 \% good agreement 
with the data is obtained \cite{Rusek05}.

In the present work we employ the standard three-body CDCC model to analyse the
elastic scattering and BU in the $^{6,7}$Li + $^{59}$Co systems. Our choice of 
systems was mainly influenced by the fact that we have already carried out 
extended CDCC calculations for both the $^{6}$Li+$^{59}$Co and $^{7}$Li+$^{59}$Co 
TF reactions \cite{Diaz03} which experimental data were previously published in 
\cite{Beck03,Souza03,Szanto03,Szanto04} and elastic scattering
data are also available for these systems \cite{Souza04}. We also present 
calculations of the TF of $^{6}$He+$^{59}$Co and $^{6}$He+$^{63,65}$Cu using 
the simplified two-body $^4$He + $^2n$ di-neutron model of $^6$He with the CDCC 
fusion model of \cite{Diaz03}. For these medium-mass targets Coulomb breakup 
effects should be smaller than with the heavy $^{209}$Bi target, so we have 
chosen not to apply the 50 \% reduction of the E1 coupling strength of 
\cite{Rusek05} here. All calculations were carried out using the code 
{\sc FRESCO} \cite{Thompson88}.

\subsection{CDCC calculations of $^{7}$Li+$^{59}$Co and
$^{6}$Li+$^{59}$Co elastic scattering} 

The CDCC calculations applied to the elastic scattering were carried out
assuming an $\alpha+d(t)$ cluster structure for $^6$Li($^7$Li). The $\alpha+d$ 
and $\alpha+t$ binding potentials were taken from refs.\ \cite{Kubo72} and 
\cite{Buck88}, respectively. However, the radius of the $\alpha+d$ binding
potential was increased to $R=2.56$ fm to obtain better agreement with the
measured $B(E2;1^+ \rightarrow 3^+)$. The $\alpha+d$ wave functions calculated
using this potential yield a $B(E2)$ of 24.0 e$^2$fm$^4$, in excellent
agreement with the measured value of $25.6 \pm 2.0$ e$^2$fm$^4$ \cite{Eig69}.
The calculations were otherwise similar to those described in 
\cite{Keeley02,Keeley03}. The continuum model space was limited to
cluster relative angular momentum values of $L$ = 0, 1, 2 and 3 for both Li 
isotopes, sufficient or more than sufficient (for the case of $^6$Li) 
to provide convergent results for the elastic scattering and BU.
The $\alpha+d(t)$ continuum was discretized into a series of bins in momentum
space of width $\Delta k = 0.2$ fm$^{-1}$ with $0.0 \leq k \leq 1.0$ fm$^{-1}$,
where $\hbar k$ denotes the momentum of the $\alpha+d(t)$ relative motion. All
couplings, including continuum--continuum couplings, up to multipolarity
$\lambda = 3$ were included. For the calculations at incident $^{6,7}$Li
energies of 18 MeV the continuum space was truncated at $k_{\mathrm{max}} =
0.8$ fm$^{-1}$. Test calculations at 30 and 26 MeV using this truncation gave
identical results to those with $k_{\mathrm{max}} = 1.0$ fm$^{-1}$. At 12 MeV
truncation of the continuum at $k_{\mathrm{max}} = 0.6$ fm$^{-1}$ was found
to be sufficient. Interaction and coupling potentials were generated using the 
cluster-folding procedure. 

The $\alpha$ + $^{59}$Co and $d(t)$ + $^{59}$Co potentials required as input 
for this procedure were obtained as follows. The $\alpha$ + $^{59}$Co
potentials were obtained by adjusting the real and imaginary well depths of the
global $\alpha$ potential of Avrigeanu {\em et al.\/} \cite{Avrigeanu94} to
match the 24.7 MeV $\alpha$ + $^{59}$Co elastic scattering data of McFadden and
Satchler~\cite{McFadden66}, resulting in normalizations $N_{\mathrm{R}} = 0.67$
and $N_{\mathrm{I}} = 2.52$ for the real and imaginary depths, respectively.
These normalizations were then applied to the global potential calculated at
the appropriate energies, there being no suitable data available to fix this
input more precisely. The $d(t)$ + $^{59}$Co potentials were the unmodified
global potentials of Perey and Perey~\cite{Perrey63} and Becchetti and
Greenlees~\cite{Becchetti71}, respectively, there being no suitable scattering
data available in the literature. 

\begin{figure}
\caption{\label{fig1}Ratios of the elastic scattering cross-sections to the
Rutherford cross sections as a function of c.m. angle for the $^{7}$Li+$^{59}$Co
system~\cite{Souza04}. The curves correspond to calculations with
(solid lines) and without (dashed lines) $^{7}$Li $\rightarrow$ $\alpha$ + $t$
breakup couplings to the continuum for incident $^7$Li energies of (a) 30 MeV,
(b) 26 MeV, (c) 18 MeV and (d) 12 MeV.}
\begin{center}
\psfig{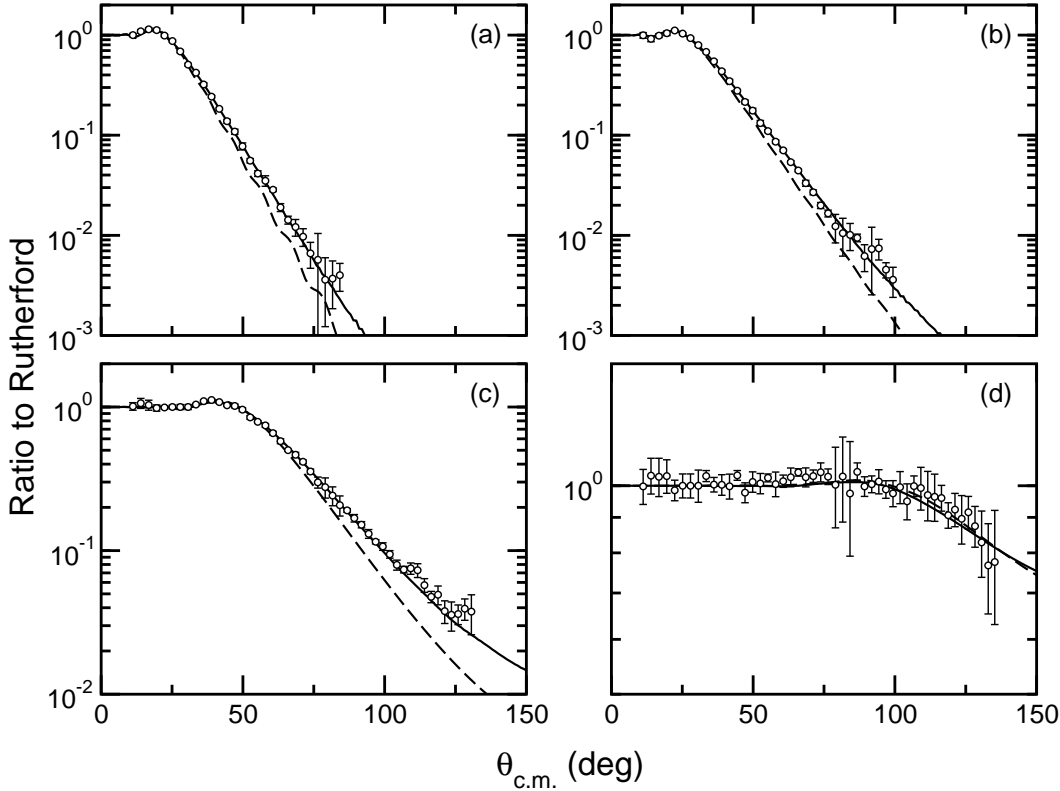}
\end{center}
\end{figure}

\begin{figure}
\caption{\label{fig2}Ratios of the elastic scattering cross-sections to the
Rutherford cross sections as a function of c.m. angle for the $^{6}$Li+$^{59}$Co
system~\cite{Souza04}. The curves correspond to calculations with
(solid lines) and without (dashed lines) $^{6}$Li $\rightarrow$ $\alpha$ + $d$
breakup couplings to the continuum for incident $^6$Li energies of (a) 30 MeV,
(b) 26 MeV, (c) 18 MeV and (d) 12 MeV.}
\begin{center}
\psfig{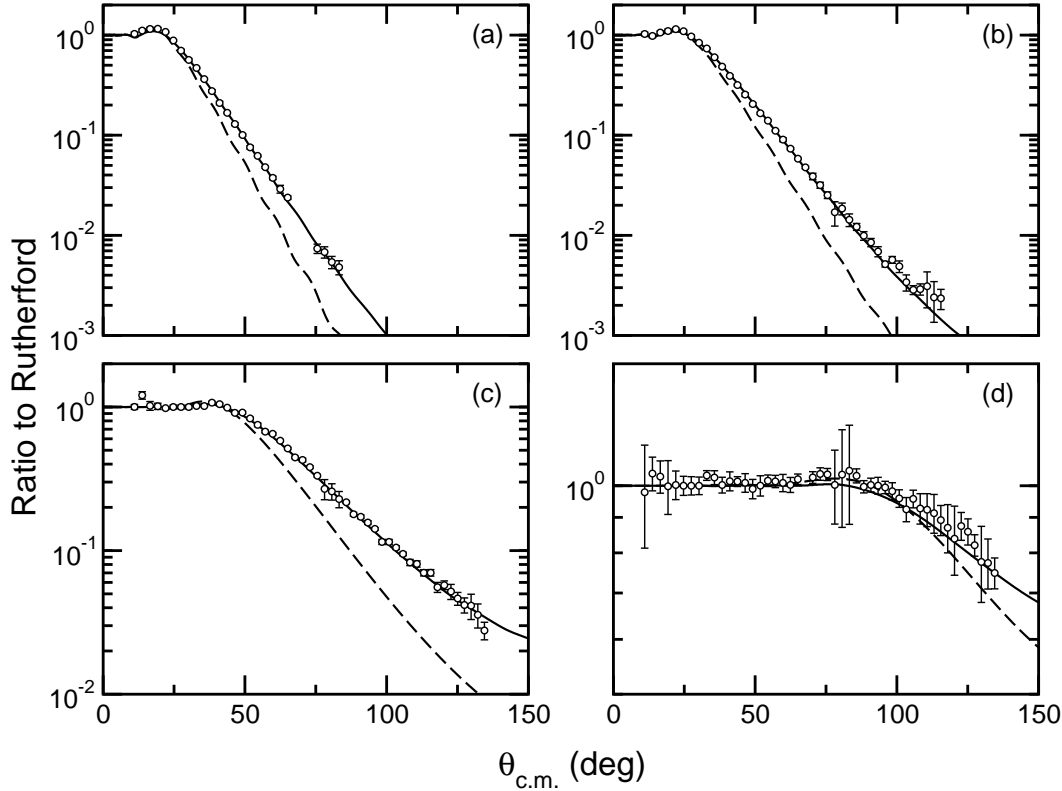}
\end{center}
\end{figure}

The real and imaginary well depths of the cluster-folded $^{6,7}$Li + $^{59}$Co
potentials (including the coupling potentials) were adjusted to obtain the
optimum description of the elastic scattering data. The CDCC calculations
are compared with the elastic scattering data of \cite{Souza04} in 
Figs.\ \ref{fig1} and \ref{fig2} for $^{7}$Li+$^{59}$Co and $^{6}$Li+$^{59}$Co, 
respectively.

The curves show the results of calculations with (solid lines) and without 
(dashed lines) $^{6,7}$Li $\rightarrow$ $\alpha$ + $d,t$ breakup couplings.
It is worth noting that the dashed line of Fig.\ \ref{fig1} has been calculated 
with reorientation of $^{7}$Li g.s. The effect of BU on the elastic scattering, 
stronger for $^{6}$Li as expected, is illustrated in Figs.\ \ref{fig1} and 
\ref{fig2} by the difference between the one-channel (i.e.\ no coupling) 
calculations and the full CDCC results. A similar effect was also observed for 
the $^{6,7}$Li+$^{65}$Cu elastic scattering~\cite{Shrivastava06} at 25 MeV 
incident energy. 

OM fits to the data were also carried out to obtain total
reaction cross sections. The starting point for the OM fits to the 
$^6$Li+$^{59}$Co data was the potential of Fulmer {\it et al.\/}
\cite{Fulmer81} for $^{6}$Li+$^{59}$Co elastic scattering at an incident 
energy of 88 MeV. For the $^7$Li + $^{59}$Co data we used the global $^7$Li 
optical potential of Cook \cite{Co82}. The real and imaginary potential depths 
and the imaginary diffuseness were searched on in both cases, all other 
parameters being held fixed. The resulting best fit parameters are given in 
Tables \ref{tab1a} and \ref{tab2a}. 

\begin{table}
\caption{\label{tab1a} OM fits to the $^7$Li + $^{59}$Co elastic
scattering data.}
\begin{tabular}{c c c c c c c c}
$E_{\mathrm{lab}}$ & $V$ (MeV) & $r_V$ (fm) & $a_V$ (fm) & $W$ (MeV) & $r_W$ (fm) & $a_W$ (fm) & $\chi^2/N$ \\
30 & 100.0 & 1.286 & 0.853 & 18.8 & 1.739 & 0.7814 & 0.88 \\
26 & 108.7 & 1.286 & 0.853 & 22.7 & 1.739 & 0.8050 & 0.64 \\
18 & 114.5 & 1.286 & 0.853 & 25.2 & 1.739 & 0.7367 & 0.52 \\
12 & 179.1 & 1.286 & 0.853 & 8.91 & 1.739 & 0.6941 & 0.66 \\
\end{tabular}
\end{table}

\begin{table}
\caption{\label{tab2a} OM fits to the $^6$Li + $^{59}$Co elastic
scattering data.}
\begin{tabular}{c c c c c c c c}
$E_{\mathrm{lab}}$ & $V$ (MeV) & $r_V$ (fm) & $a_V$ (fm) & $W$ (MeV) & $r_W$ (fm) & $a_W$ (fm) & $\chi^2/N$ \\
30 & 66.9 & 1.265 & 0.901 & 13.6 & 1.760 & 0.7632 & 1.27 \\
26 & 75.0 & 1.265 & 0.901 & 16.6 & 1.760 & 0.7675 & 0.66 \\
18 & 71.3 & 1.265 & 0.901 & 21.2 & 1.760 & 0.7905 & 1.27 \\
12 & 113.4 & 1.265 & 0.901 & 16.3 & 1.760 & 0.7114 & 0.41 \\
\end{tabular}
\end{table}

The total reaction cross sections obtained from the CDCC calculations 
are in good agreement with those obtained from the OM fits, see 
Tables \ref{tab1} and \ref{tab2}, except at 12 MeV where the relatively poor 
precision of the data means that both the OM potential parameters and the 
total reaction cross sections are poorly determined. 

We would particularly like to point out that for both systems the calculated 
total BU cross sections are negligible fractions of the total reaction 
cross sections, either calculated with CDCC or obtained from OM fits, 
which latter may be regarded as ``experimental'' values.

\begin{table}
\caption{\label{tab1}Total reaction cross sections and integrated BU
cross sections obtained from the CDCC calculations for $^7$Li + $^{59}$Co.
The total reaction cross sections extracted from OM fits to the elastic scattering
data are also given for comparison, along with the measured TF cross
sections \protect\cite{Beck03}.}
\begin{tabular}{c c c c c}
$E_{\mathrm{lab}}$ & $\sigma_R (\mathrm{OM})$ (mb) & $\sigma_R (\mathrm{CDCC})$ (mb) & $\sigma_{\mathrm{bu}}$
(mb) & $\sigma_{\mathrm{fus}}$ (mb)  \\
30 & 1603 & 1610 & 16.6 & -- \\
26 & 1547 & 1596 & 12.8 & 1014 $\pm$ 204 \\
18 & 888 & 876 & 4.67 & 547 $\pm$ 110 \\
12 & 45.4 & 83.5 & 0.31 & 38 $\pm$ 8 \\
\end{tabular}
\end{table}
 
\begin{table}
\caption{\label{tab2}Total reaction cross sections and integrated BU
cross sections obtained from the CDCC calculations for $^6$Li + $^{59}$Co.
The total reaction cross sections extracted from OM fits to the elastic scattering
data are also given for comparison, along with the measured TF cross
sections \protect\cite{Beck03}.}
\begin{tabular}{c c c c c}
$E_{\mathrm{lab}}$ & $\sigma_R (\mathrm{OM})$ (mb) & $\sigma_R (\mathrm{CDCC})$ (mb) & $\sigma_{\mathrm{bu}}$
(mb) & $\sigma_{\mathrm{fus}}$ (mb) \\
30 & 1480 & 1581 & 61.4 & -- \\
26 & 1401 & 1448 & 55.0 & 988 $\pm$ 199 \\
18 & 934 & 973 & 34.2 & 467 $\pm$ 94 \\
12 & 77.2 & 116.0 & 7.46 & 57 $\pm$ 12 \\
\end{tabular}
\end{table}

For $^6$Li+$^{59}$Co the calculated BU cross sections are between 
3.7--9.7 \% of the calculated total reaction cross sections, while for 
$^7$Li+$^{59}$Co the corresponding values are between 0.6--1.0 \%. The lower 
values for $^7$Li may be ascribed partly to the higher breakup threshold
(S$_\alpha$ = 2.47 MeV compared to 1.47 MeV for $^6$Li), partly to the presence 
of a bound excited state (the 0.478 MeV $1/2^-$) and strong ground state 
reorientation coupling, absent in $^6$Li. 

\begin{figure}
\caption{\label{fig3} CDCC calculation for the angular distribution of the 
$^6$Li $\rightarrow$ $\alpha + d$ sequential breakup via the 2.18 MeV $3^+$ state 
of $^6$Li compared to the data of Bochkarev {\em et al.\/} \cite{Bochkarev85}
as obtained for the $^{6}$Li+$^{59}$Co reaction at 41 MeV.}
\begin{center}
\psfig{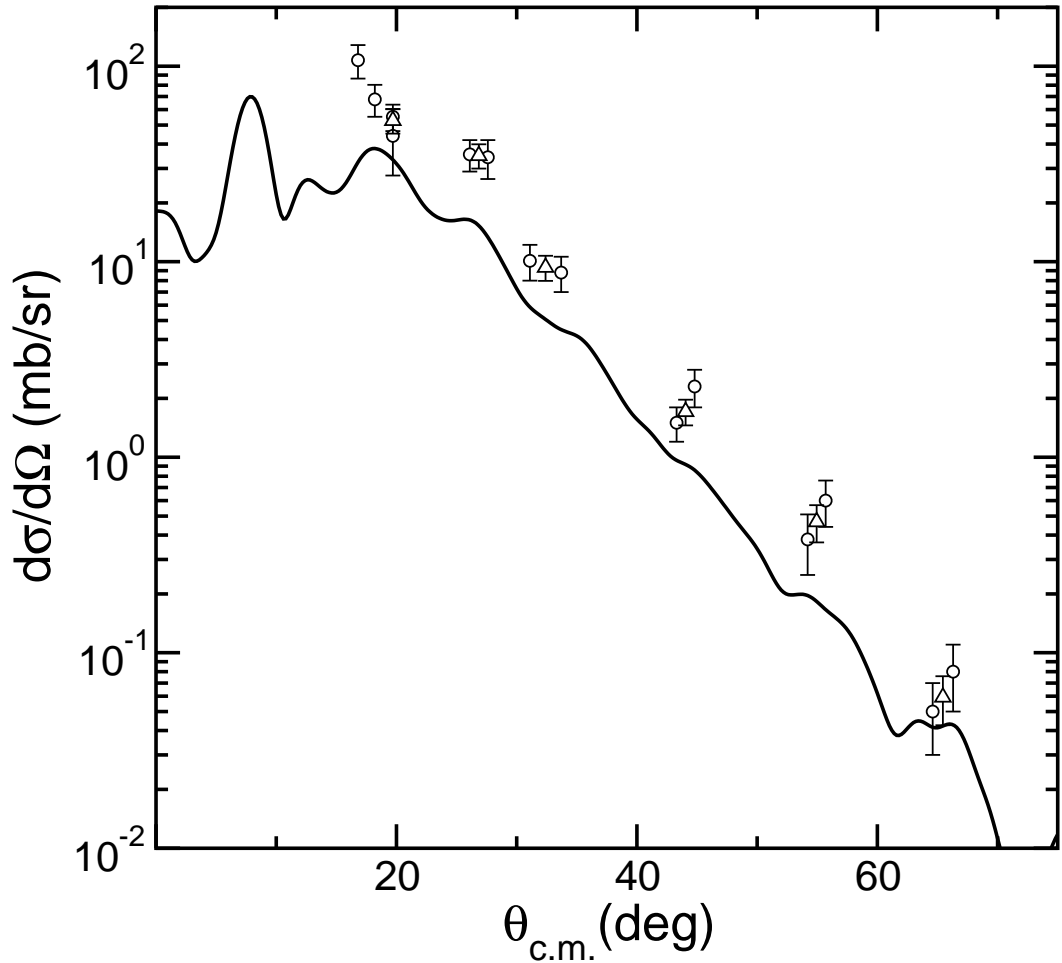}
\end{center}
\end{figure}

We may verify in part our conclusions concerning the small contribution of
BU to the total reaction cross section, as data for the sequential BU 
of $^6$Li via the 2.18 MeV $3^+$ excited state are available for a 41 MeV
$^6$Li beam incident on a $^{59}$Co target \cite{Bochkarev85}. Sequential
BU via this state is the dominant contribution to the total $^6$Li 
$\rightarrow$ $\alpha + d$ breakup cross section. There are no elastic
scattering data available at this energy, so we adjusted our CDCC calculation 
to give good agreement with the elastic scattering calculated using the
best fit OM potential parameters for the 44 MeV $^6$Li + $^{54}$Fe data of 
\cite{Kem79}, used in \cite{Bochkarev85} as the basis for a DWBA calculation 
of the ``inelastic scattering'' to the $^6$Li $3^+$ state. We compare our CDCC 
calculation with the data of \cite{Bochkarev85} in Fig.\ \ref{fig3}.
 
The calculated angular distribution is somewhat smaller than the measured one; 
a renormalization of the latter by a factor of 2/3 would result in good 
agreement with the calculation. This discrepancy in magnitude is reflected in 
the integrated cross sections; Bochkarev {\em et al.\/} \cite{Bochkarev85} 
give a value of $45 \pm 10$ mb whereas the CDCC calculation gives a value of 
22.5 mb. We note that we were unable to reproduce the data with a DWBA 
calculation using the measured $B(E2)$ value of 25.6 e$^2$fm$^4$ \cite{Eig69}
(Bochkarev {\em et al.\/} do not give details of their DWBA calculation) but 
that good agreement was obtained when we multiplied this value by 1.5. Thus, 
it is possible that there is a slight normalization factor error, of the order 
of 2/3, in the data of \cite{Bochkarev85}, in which case our calculation would 
be in excellent agreement with the data. In any case, even if the CDCC total 
BU cross sections are too small by a factor of about 1.5, this does not 
affect the conclusion that BU contributes negligibly to the total 
reaction cross section at these near-barrier energies. 

\begin{figure}
\caption{\label{fig4}Total reaction cross sections (solid curve), integrated
total BU cross sections (dot-dashed curve), integrated $^7$Li ground state
reorientation plus $1/2^-$ inelastic excitation cross sections (dashed curve)
and BPM fusion cross sections (dotted curve) obtained from the CDCC calculations
for the $^7$Li + $^{59}$Co system. The filled and open circles denote the total
reaction cross sections obtained from the best OM fits and
the measured TF cross sections \cite{Beck03}, respectively. The filled triangles
denote the summed DWBA estimates for single nucleon stripping and pickup
reactions, see text for details.}
\begin{center}
\psfig{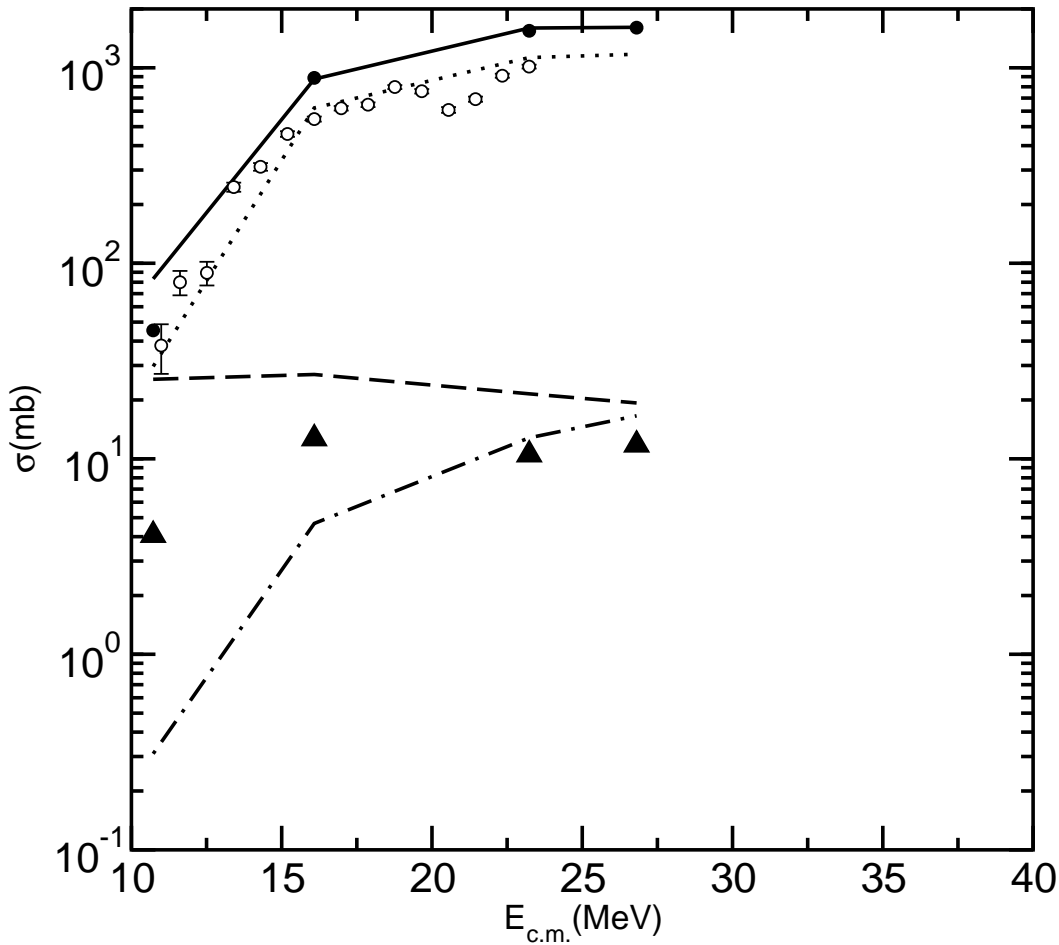}
\end{center}
\end{figure}

\begin{figure}
\caption{\label{fig5}Total reaction cross sections (solid curve), integrated
total BU cross sections (dot-dashed curve), integrated $^6$Li 2.18 MeV
$3^+$ sequential BU cross sections (dashed curve)
and BPM fusion cross sections (dotted curve) obtained from the CDCC calculations
for the $^6$Li + $^{59}$Co system. The filled and open circles denote the total
reaction cross sections obtained from the best OM fits and
the measured TF cross sections \cite{Beck03}, respectively. The open
square denotes the $^6$Li 2.18 MeV $3^+$ sequential BU cross section reported
in \cite{Bochkarev85}. The filled triangles denote the summed DWBA estimates
for single nucleon stripping and pickup reactions, see text for details.}
\begin{center}
\psfig{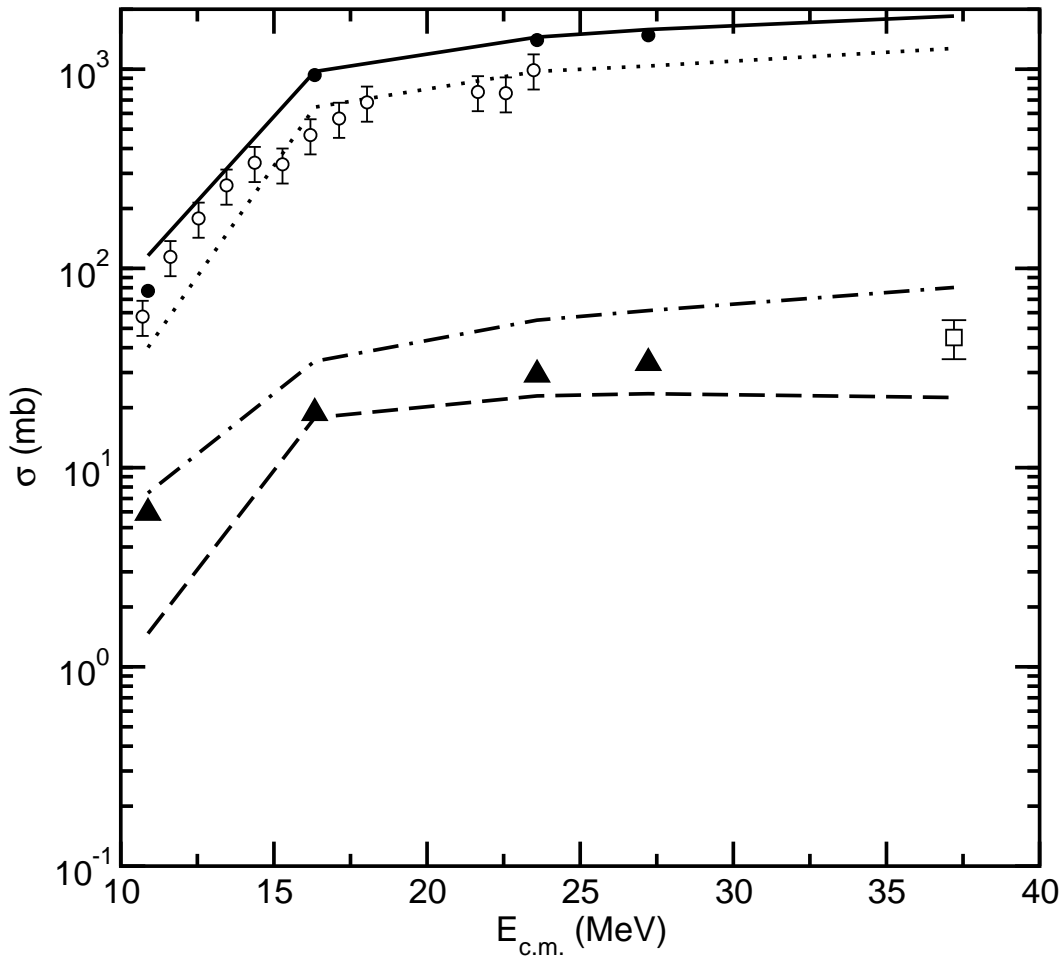}
\end{center}
\end{figure}

In Figs.\ \ref{fig4} and \ref{fig5} we present the integrated total reaction
cross sections, total BU cross sections, $^6$Li 2.18 MeV $3^+$ sequential
BU cross sections and the $^7$Li ground state reorientation plus $1/2^-$
inelastic excitation  cross sections. 

We also show the fusion cross sections obtained from barrier penetration model 
(BPM) calculations using the real part of the bare potential plus the 
``trivially equivalent local potential'' derived from the breakup couplings, the 
latter calculated using the method described in \cite{Thompson89}. These 
quantities are compared with the total reaction cross sections obtained from 
the OM fits to the elastic scattering data and the measured TF cross sections 
of \cite{Beck03}. While the method used to calculate the fusion cross sections 
is rather crude, it does appear to have some value as a means of providing a 
reasonable estimate of the TF cross section (to within about 20 \% or so) which 
may be useful when planning experiments.

It is clear from Figs.\ \ref{fig4} and \ref{fig5} and Tables \ref{tab1} and 
\ref{tab2} that the total reaction cross section is dominated by fusion at 
these near and above barrier energies (the nominal Coulomb barrier for these 
systems is equivalent to an incident Li energy of about 14 MeV). Due to the 
rather large error bars on the measured TF cross sections \cite{Beck03} it 
is not possible to draw definite conclusions, but it is evident from the tables 
that the sum of TF yields plus BU yields does not exhaust the total reaction 
cross section except for the data at 12 MeV, where the total reaction cross 
section is less well defined by the elastic scattering data. The discrepancy 
may be accounted for by inelastic excitation of the target (expected to be 
relatively unimportant for $^{59}$Co, which does not exhibit a high degree of 
collectivity), ground state reorientation plus inelastic excitation of the 
0.78 MeV $1/2^-$ state in $^7$Li and TR reactions. It should be noted 
that these other direct reactions make a considerably greater contribution to 
the total reaction cross section than does BU. The bulk of this remaining 
cross section is probably due to TR reactions of the ($^7$Li,$^6$Li), 
($^7$Li,$^8$Be), ($^6$Li,$^5$Li) etc.\ type --- as may be seen from 
Fig.\ \ref{fig4}, even when the cross sections for ground state 
reorientation and inelastic excitation of the $1/2^-$ state are added to the 
TF and BU cross sections for $^7$Li the total reaction cross 
section is far from being exhausted by the sum. 

Unfortunately, we were unable to confirm our inference by 
calculating the TR cross sections for $^{59}$Co,
as the density of states in the final nuclei is too high. However, a rough
estimate of the contribution due to single nucleon stripping and pickup
reactions was attempted through a series of DWBA calculations. Due to their
incomplete nature --- limitations in the number of
states that it was possible to include mean that the resulting cross sections are
to be regarded more as lower limits --- we give only a brief outline of the
calculations here. 

The entrance channel optical potentials were taken from Tables
\ref{tab1a} and \ref{tab2a} as appropriate, while the mass 5 and 6 and mass
7 and 8 exit channel optical potentials were calculated using the $^6$Li and
$^7$Li global parameters of \cite{Co82}, respectively. The projectile-like overlap spectroscopic
factors were taken from \cite{Co67} and the transferred nucleons were bound in Woods-Saxon
wells of radius $r_0 = 1.25$ fm and diffuseness $a = 0.65$ fm, plus a spin-orbit component
of the same geometry with a fixed depth of 6 MeV, the depth of central part being adjusted
to give the correct binding energy. The spectroscopic factors and binding potentials for
the target-like overlaps were taken from \cite{Ro78,Ba68,Ro71,Ni82}.   

The summed integrated cross sections are plotted on Figs.\ \ref{fig4} and \ref{fig5} as the filled
triangles. From  these results we may infer that the total single nucleon transfer cross sections
are at least as large as the total breakup cross sections for $^6$Li and rather larger than the
total breakup cross sections for $^7$Li over most of the incident energy range of interest here. 
Nevertheless, we are still far from being able to account for all the total reaction cross
section at the highest energies. Possible candidates for the missing cross section
are cluster transfers such as ($^6$Li,$^4$He) or ($^7$Li,$^4$He), although the large positive Q-values
for these reactions make any meaningful estimate of the cross sections impossible, as little or nothing is known of the
structure of the target-like fragments in the kinematically important excitation energy regime. 

\begin{figure}
\caption{\label{Fig:ta7}Energy dependence of the real and imaginary parts of the
bare plus DPP potentials as generated by the CDCC calculations (filled circles)
and the best OM fits potentials (open circles)
for the $^7$Li+$^{59}$Co system at a radial distance of r = 9.7 fm as
discussed in the text.}
\begin{center}
\psfig{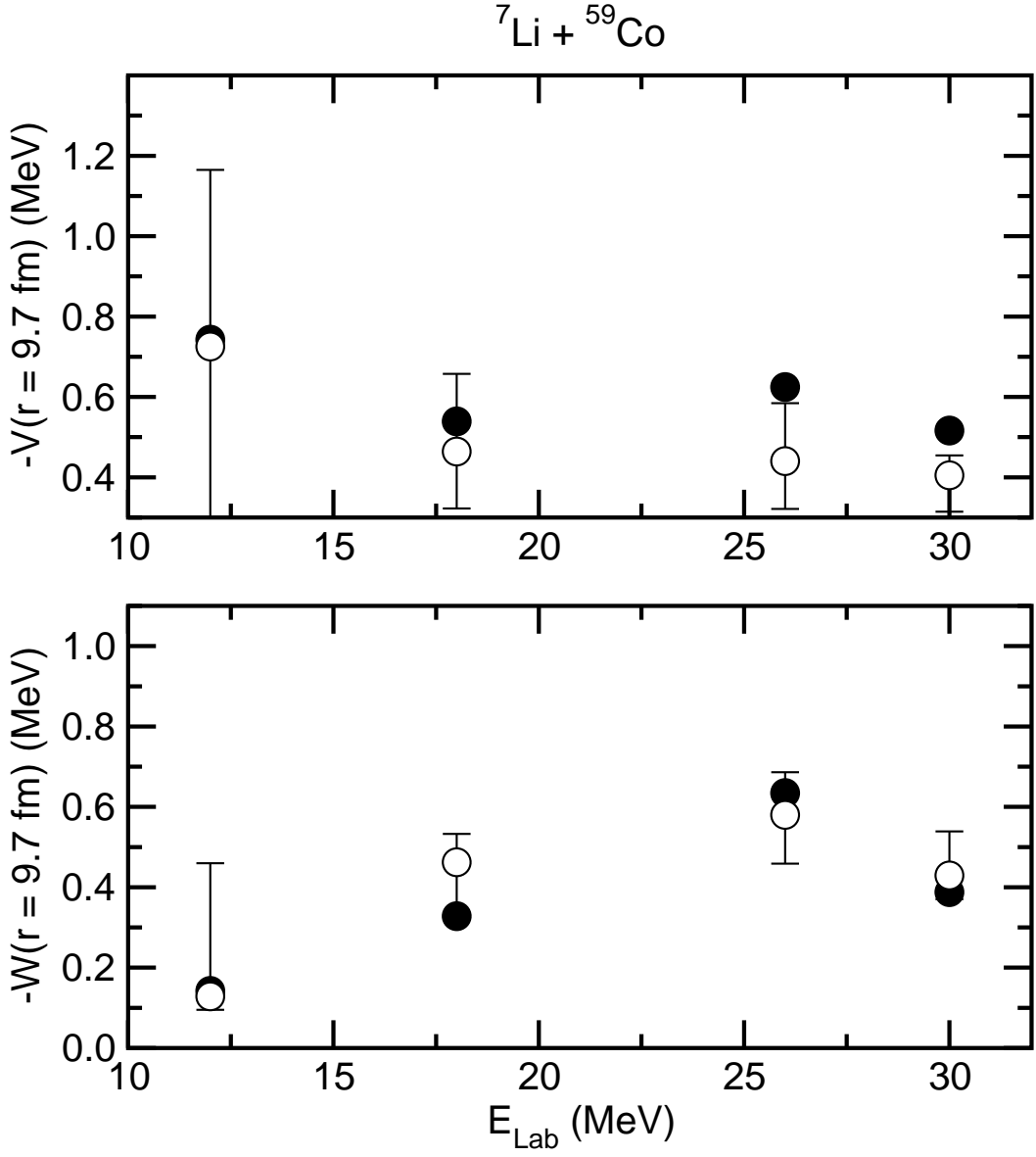}
\end{center}
\end{figure}

\begin{figure}
\caption{\label{Fig:ta6}As for Fig. \protect\ref{Fig:ta7} but for the $^6$Li+$^{59}$Co system at a radial distance
of r = 9.5 fm.}
\begin{center}
\psfig{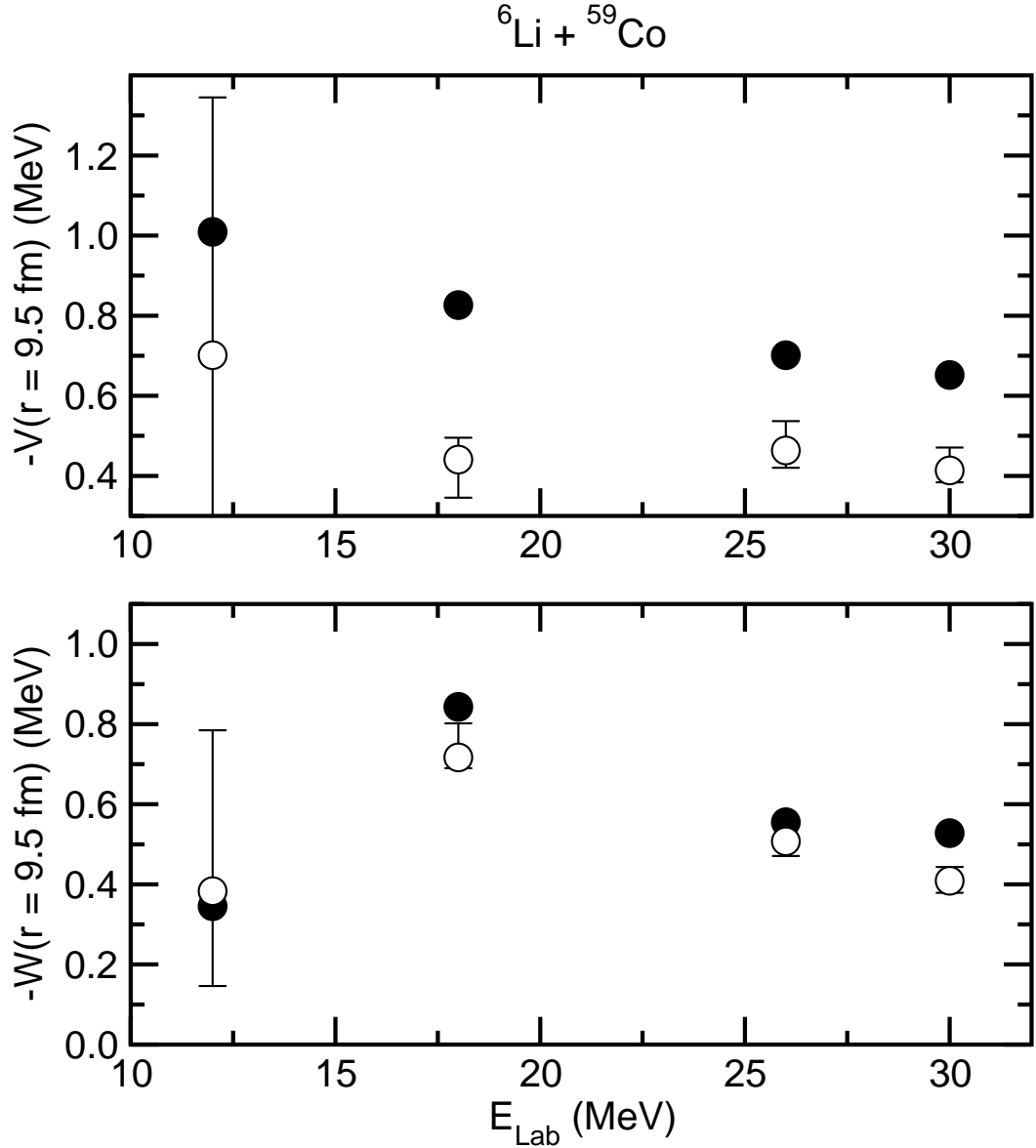}
\end{center}
\end{figure}

The real and imaginary parts of the sum of the bare plus dynamic polarization 
potentials (DPPs) generated by the couplings to BU are presented in 
Figs.\ \ref{Fig:ta7} and \ref{Fig:ta6} (filled circles) along with the best 
OM fits potentials (open circles) for comparison. The error bars on the best
fit OM values were obtained by gridding on the real diffuseness parameter
while searching on the imaginary well depth and diffuseness, all other parameters
being held fixed at the best fit values. The limits are defined by $\chi^2/N$
values of 1.0 (for those data sets where the best fit $\chi^2/N$ value is less
than 1.0) or 15 \% larger than the minimum value (for those data sets where
the minimum $\chi^2/N$ is greater than 1.0).

The potentials are evaluated at radial distances of 9.7 and 9.5 fm for $^7$Li 
and $^6$Li, respectively. These values are the mean strong absorption radii 
obtained from the best OM fits potentials at 18, 26 and 30 MeV for each system 
(the results at 12 MeV were omitted due to the large uncertainties in the 
OM fits to these data). The difference of 0.2 fm in the ``radii of 
sensitivity'' is not significant, as in reality the elastic scattering data 
probe the nuclear potential over a region of width $\sim 1$ fm in the nuclear 
surface at a given energy rather than at a single radius (which latter, if 
taken at face value, would violate the uncertainty principle, see e.g.\ 
\cite{Mac81}).

In general, the surface strengths of the ``bare plus DPP'' potentials are in 
very good agreement with those of the best OM fits  potentials, the 
exception being the real potentials for the $^6$Li+$^{59}$Co system where the 
total potentials derived from the CDCC calculations are consistently larger 
than the OM values. At first sight, one would conclude that the 
surface potential strengths for both systems exhibit the energy dependence 
characteristic of the ``threshold anomaly'', i.e.\ a rise in the strength of 
the real part as the incident energy is reduced towards the Coulomb barrier 
accompanied by a drop in that of the imaginary part. However, this conclusion 
largely rests on the values at 12 MeV incident energy, and as can be seen
from the error bars, the potential strength in the nuclear surface is 
effectively not determined by the data due to its rather poor precision, a 
very wide range of values giving equally good fits to the data for both 
systems. The spread in values for the other energies, while much less than 
that at 12 MeV, is still such that we are unable to draw any concrete 
conclusions concerning the presence or absence of a threshold anomaly (TA) 
in these systems. 

\subsection{CDCC calculations of $^{6}$He+$^{59}$Co fusion reaction}

Calculations employing the CDCC model for TF of \cite{Diaz03} were 
also carried out to describe the fusion process induced by the ``Borromean'' 
nucleus $^{6}$He on the same medium-mass target $^{59}$Co. Firstly, we would 
like to stress that in these calculations --- unlike those for the $^{6,7}$Li 
elastic scattering and BU reported in the previous section --- the 
imaginary components of the off-diagonal couplings in the transition 
potentials were neglected, while the diagonal couplings include imaginary 
parts \cite{Diaz03}. Otherwise full continuum couplings were taken into 
account. We used short-range imaginary potentials for each projectile fragment 
separately (for example, $\alpha$ and $d$+target potentials for the case of 
the $^{6}$Li nucleus). This is equivalent to the use of the incoming wave 
boundary condition in CCFULL calculations~\cite{Beck03}; however, only the TF
cross sections can be evaluated with this model. Ideally, one would wish to 
employ this version of CDCC in a single calculation that attempts to describe 
the ensemble of the data, TF, BU, TR and elastic scattering. However, we are 
still some way from being able to include all the necessary direct reaction 
couplings in a single practicable calculation, at least for systems where 
fusion data exist (this problem applies equally well to the stable weakly 
bound nuclei as well as $^6$He).

The calculations were similar to those described in more detail in 
\cite{Diaz03} for $^{6}$Li, but now applied to the two-neutron halo nucleus 
$^{6}$He. The present case is much more complicated since $^{6}$He breaks 
into three fragments ($\alpha$+n+n) instead of two ($\alpha$+d), and the CDCC 
method for two-nucleon halo nuclei has not yet been implemented in FRESCO. 
Hence a di-neutron model is adopted for the $^{6}$He+$^{59}$Co reaction, 
i.e.\ we assume a two-body cluster structure of $^{6}$He = $^{4}$He+$^{2}$n 
with an $\alpha$ particle core coupled to a single particle representing a 
di-neutron ($^{2}$n) like cluster.

\begin{figure}
\caption{\label{Fig:8} Energy dependence of TF cross sections for the 
$^6$He+$^{59}$Co reaction obtained with the CDCC method~\cite{Diaz03}. The 
dashed and thin curves correspond respectively to CDCC calculations with and 
without continuum couplings. The experimental TF cross sections for the 
$^6$Li+$^{59}$Co system~\cite{Beck03} are given for the sake of comparison. For 
each reaction, the incident energy is normalized by the Coulomb barrier of the 
effective potential~\cite{Broglia91,Christensen76}.}
\begin{center}
\psfig{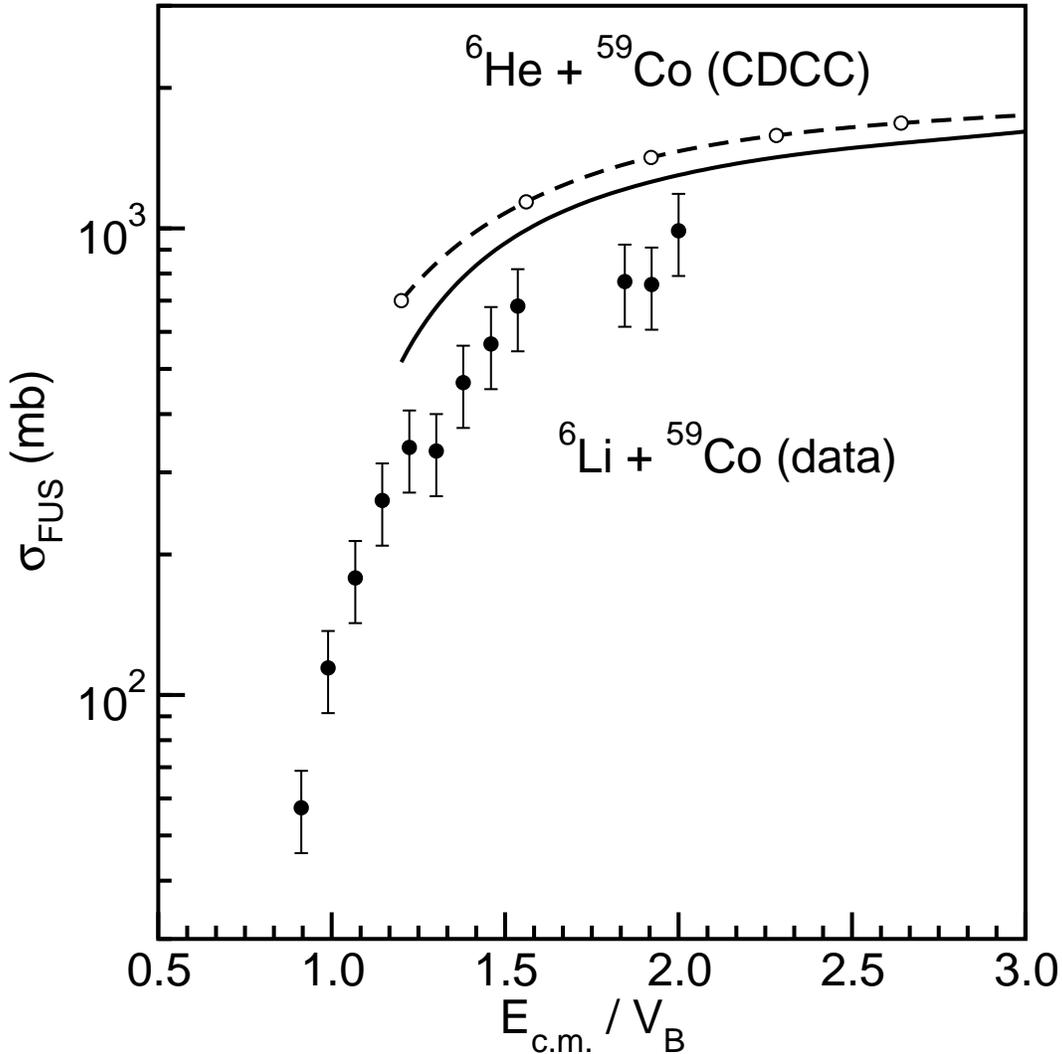}
\end{center}
\end{figure}

\begin{figure}
\caption{\label{Fig:9} The $^6$He+$^{59}$Co TF excitation functions 
are the same as in Fig.~8 and are compared with $^4$He+$^{59}$Co TF excitation 
functions. The TF cross sections of $^4$He+$^{59}$Co were taken from 
\cite{McMahan80} and standard calculations (solid curve) were performed 
as discussed in the text.}
\begin{center}
\psfig{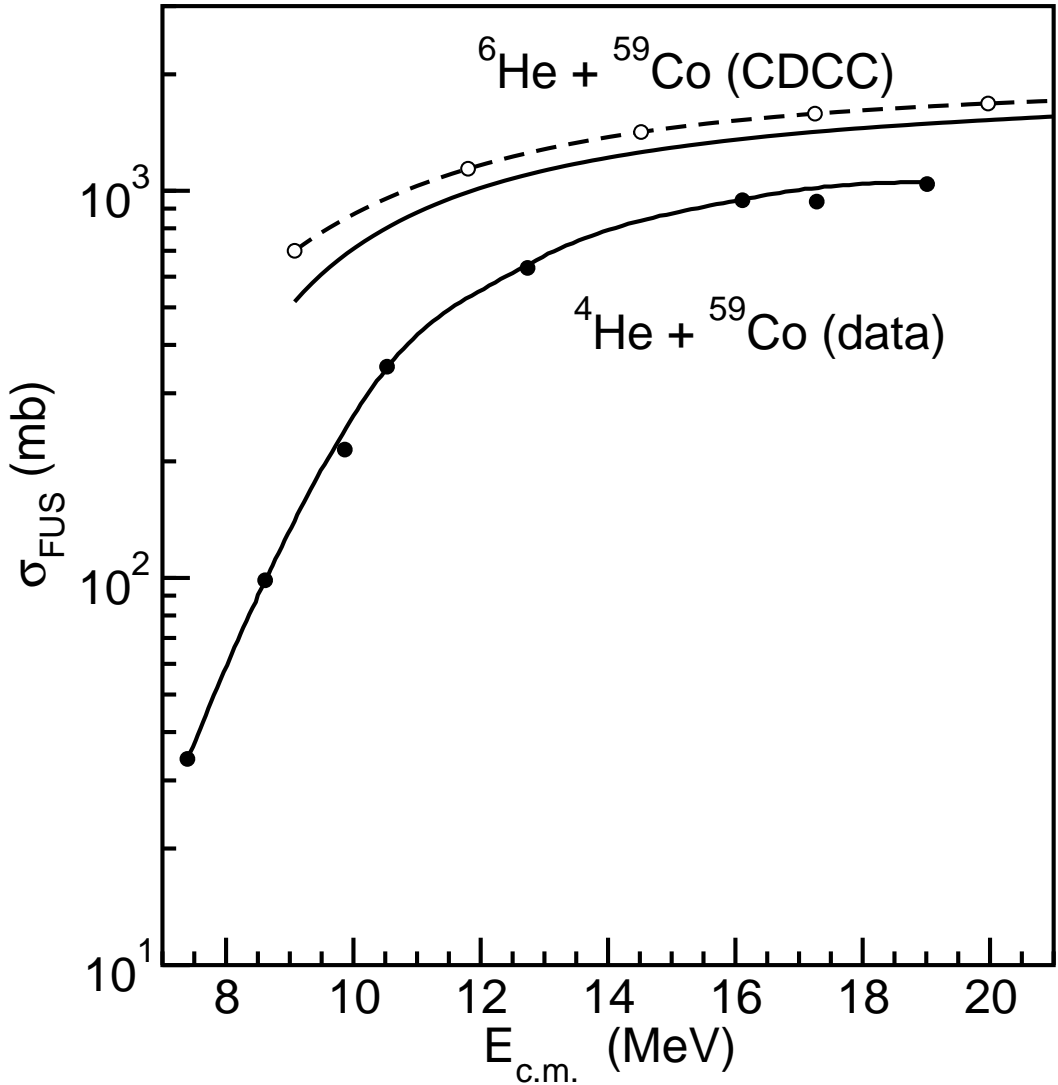}
\end{center}
\end{figure} 
 
As in our previous work~\cite{Diaz03}, the real part of the potentials 
between the fragments and the $^{59}$Co target are those obtained with the global 
Broglia-Winther Woods-Saxon parametrization~\cite{Broglia91,Christensen76}. The 
numerical values for $^{2}$n-$^{59}$Co and for $\alpha$-$^{59}$Co are: 
V$_{o}$ = -16.89 (-31.14) MeV, r$_{o}$ = 1.09 (1.127) fm and a = 0.63 (0.63) fm. 
For the $\alpha$-2n binding potential (0$^{+}$ g.s.) we have used the following
Woods-Saxon potential: V$_{o}$ =  -40.796 MeV, r$_{o}$ = 1.896 fm and a = 0.3
fm. The g.s. binding potential of the $\alpha$ particle and the di-neutron
provides a 2s bound state of about -0.975 MeV. The binding potential of the
2$^{+}$ resonant state also has a Woods-Saxon form with the following parameters: 
V$_{o}$ = -35.137 MeV, r$_{o}$ = 1.896 fm, a = 0.3 fm. With this potential the 
energy of the 2$^{+}$ resonant state in $^{6}$He is 0.826 MeV and its width is 
0.075 MeV. To obtain converged (within a 5$\%$ level) TF cross section we have 
included: (i) couplings to the 2$^{+}$ resonant state and non-resonant continuum 
(BU) states with partial waves for $\alpha$-2n relative motion up to f-waves 
($L$ = 0, 1, 2, and 3), (ii) the $^{6}$He fragment-target potential multipoles 
up to the octopole term, and (iii) a maximum continuum energy of 8 MeV. All 
continuum couplings (including both bound--continuum couplings and 
continuum--continuum couplings) were included in the calculation. 

\begin{table}
\caption{\label{tab3} TF cross sections obtained from the CDCC calculations
for $^6$He + $^{65}$Cu without (th$_1$) and with (th$_2$) continuum couplings are compared
with measured total residue cross sections (exp) \protect\cite{Navin04}. The total reaction
cross sections extracted from OM fits to the elastic scattering data and deduced BU cross 
sections \protect\cite{Navin04} are also given for comparison.}
\begin{tabular}{c c c c c c}
$E_{\mathrm{lab}}$ (MeV) & $\sigma_{th_1}$ (mb) & $\sigma_{th_2}$ (mb) & $\sigma_{\mathrm exp}$ (mb) & $\sigma_R (\mathrm{OM})$ (mb) & $\sigma_{\mathrm bu}$ (mb)  \\
30 & 1637 & 1846 & 1334 & 1614 & 280 \\
19.5 & 1371 & 1606 & 1292 & 1502 & 210 \\
\end{tabular}
\end{table} 

\begin{table}
\caption{\label{tab4} TF cross sections obtained from the CDCC calculations
for $^6$He + $^{63}$Cu without (th$_1$) and with (th$_2$) continuum couplings are compared
with total residue cross sections (exp) measured at 30 MeV \protect\cite{Navin04}. The total 
reaction cross sections and BU cross sections were not reported in \protect\cite{Navin04}.}
\begin{tabular}{c c c c c c}
$E_{\mathrm{lab}}$ (MeV) & $\sigma_{th_1}$ (mb) & $\sigma_{th_2}$ (mb) & $\sigma_{\mathrm exp}$ (mb) & $\sigma_R (\mathrm{OM})$ (mb) & $\sigma_{\mathrm bu}$ (mb)  \\
30 & 1600 & 1830 & 1400 & -- & -- \\
19.5 & 1349 & 1585 & -- & -- & -- \\
\end{tabular}
\end{table}

Results of the CDCC calculations for the TF fusion of $^{6}$He+$^{59}$Co 
system are compared in two ways. First we present a comparison with the 
experimental excitation function of the $^{6}$Li+$^{59}$Co system \cite{Beck03} 
as displayed in Fig.\ \ref{Fig:8}. An equivalent comparison with 
$^{4}$He+$^{59}$Co data \cite{McMahan80} is given in Fig.\ \ref{Fig:9}. Note that 
the calculation presented for the latter system is a simple two-body scattering 
calculation with an OM potential with an interior imaginary part simulating the 
ingoing wave boundary condition. In both cases we note that the bare no coupling 
TF calculation is already considerably larger than the TF cross sections for 
either $^{6}$Li+$^{59}$Co or $^{4}$He+$^{59}$Co, and that the breakup couplings 
further increase the calculated TF cross sections at all energies investigated 
here.

Calculations were also performed for other medium-mass targets such as 
$^{63,65}$Cu and $^{64}$Zn nuclei. Their results are summarized for the copper 
isotopes in Tables V and VI along with experimental results reported in 
\cite{Navin04}. Here we again see that the effect of the breakup couplings is to 
increase the TF fusion cross section. However, the final values are rather 
larger than the data \cite{Navin04}, of the order of 20--30 \%. A simlilar 
conclusions is found for the zinc target at both near-barrier and sub-barrier
energies \cite{Dipietro04}.  This discrepancy could be due to the real potentials 
used (particularly when used for the $^2$n + target potentials) given that the 
bare no coupling values for the TF are already slightly larger than the measured 
ones or may be indicative of other coupling effects; coupling to single neutron 
stripping has been found to significantly reduce the TF cross section for $^6$He 
at similar energies with respect to the Coulomb barrier \cite{Keeley07}.

\section{Discussion} 

It has already been remarked that there is some confusion about the definition 
of fusion~\cite{Canto06,Diaz03}. Theorists usually define complete fusion (CF) 
as the capture of all the projectile fragments, and incomplete fusion (ICF) as 
the capture of only some fragments~\cite{Diaz03}. As in all other CC 
calculations, CDCC has the disadvantage of being unable to distinguish between 
CF and ICF. The combined effect of BU and TR in the CC approach has not been 
studied so far in the context of sub-barrier fusion. Another complication in 
experiments arises from a clear separation of CF and ICF cross sections, 
therefore CF is often defined experimentally as the capture of all the charge of 
the projectile by the target \cite{Dasgupta04,Liu05}, although this definition 
would lead to problems for neutron halo nuclei such as $^6$He. In the following 
we discuss only TF cross sections (the sum of CF and ICF cross sections).

The standard three-body CDCC model is adequate for $^{6,7}$Li as core excitation 
may be safely ignored for an $\alpha$ particle core. The elastic scattering 
data \cite{Souza04} as plotted in Figs.\ \ref{fig1} and \ref{fig2} are found to 
be very well reproduced for both the $^{7}$Li and $^{6}$Li nuclei, at least for 
the three highest incident energies. It is clear that despite the essentially 
negligible contribution of BU to the total reaction cross section coupling
to BU has an important effect on the elastic scattering for both systems.  
Although the total reaction cross sections are dominated by TF, it is also clear 
that the sum of TF+BU by no means exhausts the total reaction cross section. As 
target excitation is expected to be relatively weak for $^{59}$Co this leaves 
TR reactions as the other main contributor to the total reaction cross section, 
see e.g.\ \cite{Shrivastava06} and \cite{Pakou03a,Pakou05} for medium-mass and 
light targets, respectively. The effect of TR coupling on elastic scattering for 
weakly bound light projectiles remains to be fully elucidated, although it could 
be important depending on the system, see e.g.\ \cite{Keeley97,Keeley04a}. It has 
already been demonstrated that the form of CDCC adapted to TF calculations is 
able to well describe TF for the $^{6,7}$Li + $^{59}$Co systems \cite{Diaz03}.

Less clear is the question of whether either system exhibits a TA. Within the 
uncertainties, the surface strengths of the real parts of the best fit OM 
potentials show no dependence on incident energy for either isotope. This is also 
true for the imaginary part for $^7$Li, while the imaginary part for $^6$Li seems 
to show a gradual rise in surface strength as the incident energy is reduced 
towards threshold, as seen previously for other targets \cite{Keeley94,Pakou03b}. 
However, given the somewhat artificial constraints employed in the grid searches 
carried out to define the error bars on the OM potentials one may equally argue 
that the $^6$Li imaginary potential surface strength is also consistent with 
little or no variation with incident energy.

An overview of the available elastic scattering data for lithium isotopes from a 
variety of targets: $^{208}$Pb \cite{Keeley94}, $^{138}$Ba \cite{Maciel99}, 
$^{28}$Si \cite{Pakou04}, and $^{27}$Al \cite{Figueira06,Figueira07} does not allow 
any firm general conclusions concerning the presence or absence of the TA for 
either isotope. Part of the problem lies in the need for high precision data if one 
is to reduce the ambiguities in the extracted OM potential surface strengths to a 
level where firm conclusions as to their dependence on incident energy may be 
drawn. This is particularly true for the region around the ``Coulomb rainbow'' for 
the real part of the potential. There is also the question of dependence on 
target mass; there is no {\em a priori\/} reason to suppose that the TA found to 
be present in the $^7$Li+$^{208}$Pb system \cite{Keeley94} will necessarily also 
be present in a system with a lighter target. For the $^7$Li+$^{208}$Pb system it 
was shown that coupling to the $^{208}$Pb($^7$Li,$^6$Li)$^{209}$Pb transfer, with 
a negative reaction Q-value, could account for the presence of TA \cite{Keeley97}. 
However, for a $^{58}$Ni target the reaction Q-value for the same stripping 
reaction is now positive, and it has been found that TR reactions with positive 
Q-values can give rise to DPPs that have similar properties to those produced by 
BU couplings \cite{Keeley04a}. From the present analysis with the $^{59}$Co
medium-mass target, it still remains unclear how the BU coupling affects the TA 
present for all tightly bound nuclei and if the concept of BU threshold anomaly 
\cite{Figueira07,Hussein06} will be needed.

With no data available for $^{6}$He+$^{59}$Co we cautiously decide not to 
present CDCC calculations for the elastic scattering for this system as the 
Coulomb dipole excitation is known to be too strong in the di-neutron 
approximation \cite{Rusek05}, although a similar core--di-neutron 
model~\cite{Esbensen97} is capable of describing reasonably well the main 
properties of $^{6}$He; e.g.\ the nuclear charge radius, which measurements 
recently reported with high precision~\cite{Wang04}, was well predicted (to 
within 5 \%). The dipole Coulomb excitation of $^{6}$He projectiles in the field 
of a highly charged target has already been discussed 
\cite{Rusek03,Keeley03,Mackintosh04,Rusek05}. The di-neutron CDCC model has been 
found to give much better agreement with elastic scattering data when the dipole 
coupling strength is reduced by 50 $\%$ \cite{Rusek05}. This reduction is 
important for heavy targets, but probably not as much for a medium-mass target
like $^{59}$Co. Nevertheless, such a reduction also reduces the total 
absorption cross section in the CDCC calculations. If we consider this cross 
section as approximating to the TF cross section, we may overestimate the 
fusion of $^{6}$He+$^{59}$Co slightly. 

The CDCC calculations for the $^6$He + $^{59}$Co TF described in the previous 
section are displayed in Figs.\ \ref{Fig:8} and \ref{Fig:9}. They do not include 
either target excitations or TR channels. However, crude estimations such as 
those performed for the $^{6}$Li+$^{59}$Co reaction \cite{Diaz03} find the effect 
of target excitation to be very small. In Fig.\ \ref{Fig:8} we compare the TF 
excitation functions for $^{6}$He+$^{59}$Co (CDCC calculations) and 
$^{6}$Li+$^{59}$Co (experimental data of \cite{Beck03}). We note that both 
calculated curves for $^6$He, with (dashed line) and without (solid line) BU 
couplings, give much larger TF cross sections than for $^{6}$Li. Similar 
conclusions are reached when the $^{6}$He+$^{59}$Co TF excitation function (CDCC 
calculations) is compared to that for $^{4}$He+$^{59}$Co (here standard 
calculations fit the data of \cite{McMahan80} remarkably well) in 
Fig.\ \ref{Fig:9}. This is a general result for medium-mass targets and does not 
depend on the nature of the target, as shown in Tables V and VI for two 
different copper isotopes. However, the calculations for $^6$He + $^{63,65}$Cu 
somewhat over predict the measured TF cross sections. This could be due to the 
bare potentials used as input (the bare no coupling calculations give TF cross 
sections that are larger than the measured values, and the BU coupling 
consistently leads to an increase of the TF cross section in the CDCC model), 
the overestimation of the BU coupling effect on TF due to the use of the 
two-body di-neutron model of $^6$He, or the effect of TR couplings, found 
to decrease the TF cross section for $^6$He at similar incident energies with 
respect to the Coulomb barrier \cite{Keeley07}. Unfortunately, this latter 
hypothesis cannot be tested in these systems due to the high density of states 
in the residual nuclei involved, ruling out a practicable calculation. 

The present CDCC results, i.e.\ an increase in the TF cross section due to BU 
couplings, are in agreement with an alternate CC approach proposed by Dasso and 
Vitturi~\cite{Dasso05} that mimics continuum--continuum couplings in the BU 
channels. However, contradictory results have been obtained by Ito 
{\em et al.\/} \cite{Ito06} with a different approach based on a time-dependent 
wave-packet formalism suggesting the possible importance of higher partial waves 
for the relative motion between the valence particle and the projectile core. 
The converged cross sections within the CDCC approach (the study of the 
convergence of the results with respect to the number of angular momentum states 
in the continuum is discussed with great care in \cite{Diaz03}) are found to be 
in reasonable agreement with the available TF data for medium-mass 
targets~\cite{Navin04} (see Table III). This conclusion is consistent with 
similar CDCC calculations performed for heavy targets
\cite{Rusek03,Rusek04,Rusek05} and using the di-neutron model. It should be 
mentioned that a recent study \cite{Matsumoto06} of the $^{6}$He+$^{209}$Bi 
reaction indicates that the $\alpha$+$n$+$n$+$^{209}$Bi four-body model provides 
a more accurate description of the $^6$He elastic scattering within the CDCC 
formalism than the di-neutron model. It would be interesting to see what 
difference this more accurate model would have on the BU coupling effect on TF 
if applied to a fusion calculation in a similar manner to the calculations 
presented here.

\section{Summary and concluding remarks} 

Halo and weakly bound cluster nuclei are good test-benches for theories of BU 
and fusion. We have shown that strong CC effects can be taken into account within 
a CDCC approach to model breakup effects on the angular distributions of the 
elastic scattering and on the excitation functions of the total (CF + ICF) fusion 
for reactions induced by $^{6,7}$Li and $^{6}$He projectiles. Although BU does 
not contribute significantly to the total reaction cross section at near-barrier 
energies, its influence is decisive for a fairly good description of the 
$^{6,7}$Li+$^{59}$Co elastic scattering data. For both systems the total reaction 
cross sections are dominated by fusion at near and above barrier energies. The 
CDCC calculations suggest that there are other direct reaction processes (most 
likely nucleon TR) with larger contributions to the total reaction cross section 
than BU. The real and imaginary parts of the $^{6}$Li+$^{59}$Co and 
$^{7}$Li+$^{59}$Co DPPs generated from the best OM fits to their respective 
elastic scattering angular distributions do not allow us to draw any concrete 
conclusions concerning the occurrence or not of the TA phenomenon in these 
systems.

Near-barrier TF cross sections calculated by CDCC for $^{6}$He+$^{59}$Co 
are much larger than the measured TF yields for $^{6}$Li+$^{59}$Co 
\cite{Beck03} and $^{4}$He+$^{59}$Co \cite{McMahan80} that are well reproduced 
by calculations. However, similar CDCC calculations for the 
$^6$He+$^{63,65}$Cu systems somewhat over predict the data \cite{Navin04}.
This may be due to deficiencies in the two-body model of $^6$He used, to the
global potentials used as a basis for the calculations or to the neglect of
coupling to other reaction channels, e.g.\ TR. The present CDCC calculations
show a consistent enhancement of the TF cross section due to coupling to BU
with respect to the no coupling calculations. However, for a general description 
of fusion induced by $^{6}$He projectiles a more complete theoretical model of 
few-body quantum dynamics that is able to: i) distinguish CF from ICF and ii)
treat explicitly TR channels is required and which will need to follow 
correlations after BU \cite{Diaz02}. The two-body CDCC calculations
\cite{Diaz02,Keeley02,Rusek03,Diaz03,Keeley03,Rusek04,Rusek05} of the type we 
have presented in this work can serve as a good starting point; while $^{6}$He 
is best described as a three-body $\alpha$-$n$-$n$ object, a two-body 
$\alpha$-$^{2}$n model appears to be satisfactory provided the E$_{1}$ strength 
is reduced by a factor of $\sim 0.5$ \cite{Rusek05}. This is consistent with the 
fact that the mean charge radius measured for the two-neutron halo nucleus 
$^{6}$He~\cite{Wang04} can be fairly well described by di-neutron cluster
models~\cite{Esbensen97}. 

The CDCC formalism, with continuum--continuum couplings taken into account, is 
probably one of the most reliable methods available nowadays to study reactions 
induced by weakly bound nuclei and exotic halo nuclei, although many of the
latter have added complications like core excitation and three-body structure 
that are only now being incorporated within the formalism 
\cite{Matsumoto04a,Matsumoto04b,Rodriguez05,Matsumoto06,Summers06a,Summers06b}. 
One really needs to investigate such processes within the dynamics of the 
interaction at the Coulomb barrier with loosely bound halo nuclei. An 
understanding of the reaction dynamics involving couplings to BU channels 
requires the explicit measurement of elastic scattering data with a high degree 
of precision as well as yields leading to the TR and BU channels. The complexity 
of such reactions, where many processes compete on an equal footing, necessitates 
kinematically and spectroscopically complete measurements, i.e.\ ones in which 
all processes from elastic scattering to fusion are measured simultaneously, 
providing a technical challenge in the design of broad range detection systems. 
A systematic study of $^{6}$He induced fusion reactions with the CDCC method is 
still awaited, as up to now only very few experimental studies with $^{6}$He 
projectiles are available 
\cite{Kolata98,Raabe04,Bychowski04,Dipietro04,Navin04,Penion06}. A new 
experimental program with SPIRAL beams and medium-mass targets is getting 
underway at GANIL. 
 
The application of four-body CDCC models under current development
\cite{Matsumoto04a,Matsumoto04b,Rodriguez05,Matsumoto06,Summers06a,Summers06b} 
will then be highly desirable. The questions in the theory of a two-neutron 
halo system such as $^{6}$He, its BU (and in the breakup of many-body 
projectiles generally), and its CF and ICF will need knowledge not just of 
those integrated cross sections, but the phase space distributions of the 
surviving fragment(s). Therefore, future very exclusive experiments will need 
to determine very precisely the spatial (angular and energy) correlations of 
the individual neutrons. Preliminary attempts at measurements 
\cite{Bychowski04,DeYoung05} of $\alpha$-particles in coincidence with neutrons 
are promising. 

\section{Acknowledgments} 

The authors would like to warmly thank F.~A. Souza for supplying the elastic
scattering cross sections of \cite{Souza04} in tabulated form. We also 
acknowledge P.~R.~S. Gomes, S. Kailas, R. Raabe, K. Rusek, and V. Zagrebaev 
for stimulating and instructive discussions. One of us (N.K.) gratefully 
acknowledges the receipt of a Marie Curie Intra-European Fellowship from the
European Commission, contract No. MEIF-CT-2005-010158.

\end{document}